\documentclass[a4paper,10pt]{article}

\usepackage[
  top=15mm, 
  bottom=15mm, 
  left=20mm, 
  right=20mm,
  heightrounded]{geometry}

\usepackage[utf8]{inputenc}
\usepackage[dvips]{graphicx}
\usepackage{amssymb}
\usepackage{latexsym}
\usepackage{amsmath}
\usepackage{lineno}
\usepackage{color}
\usepackage{xspace}

\usepackage{longtable} 
\usepackage{rotating}
\usepackage{multirow}
\usepackage{setspace}

\usepackage{todonotes}

\newcommand{\ie}{i.e.,\xspace}
\newcommand{\eg}{e.g.,\xspace}

\usepackage{authblk}

\title{Evolutionary dynamics of organised crime and terrorist networks}
\author[1,2]{Luis A. Martinez-Vaquero} 
\author[3,4]{Valerio Dolci}
\author[1]{Vito Trianni} 
\affil[1]{Institute of Cognitive Sciences and Technologies, 
National Research Council of Italy,\newline via San Martino della Battaglia 44,
00185 Rome, Italy}
\affil[2]{\textit{Current address:} Lab of Socioecology and Social Evolution, Department of Biology, KU Leuven,
Naamsestraat 59, 
3000 Leuven, Belgium}
\affil[3]{INFN Roma1, Rome, Italy}
\affil[4]{Physics Department, Sapienza University of Rome, Italy}

\date{}

\begin{document}

\maketitle

\begin{tabular}{l}
\textbf{Corresponding author}\\
Dr. Luis A. Martinez-Vaquero\\
Department of Biology, KU Leuven\\
Naamsestraat 59\\ 
3000 Leuven, Belgium\\
email: l.martinez.vaquero@gmail.com 
\end{tabular}

\begin{abstract}
  Crime is pervasive into modern societies, although with different
  levels of diffusion across regions.  Its dynamics are dependent on
  various socio-economic factors that make the overall picture
  particularly complex. While several theories have been proposed to
  account for the establishment of criminal behaviour, from a
  modelling perspective organised crime and terrorist networks
  received much less attention. In particular, the dynamics of
  recruitment into such organisations deserve specific considerations,
  as recruitment is the mechanism that makes crime and terror
  proliferate. We propose a framework able to model such processes in
  both organised crime and terrorist networks from an evolutionary
  game theoretical perspective. By means of a stylised model, we are
  able to study a variety of different circumstances and factors
  influencing the growth or decline of criminal organisations and
  terrorist networks, and observe the convoluted interplay between
  agents that decide to get associated to illicit groups, criminals
  that prefer to act on their own, and the rest of the civil society.
\end{abstract}

\section{Introduction}
\label{sec:introduction}

Criminal organisations (COs) and terrorist networks (TNs) represent
different outcomes of a similar process, whereby individuals join
together into illicit groups to bring forth criminal activities at the
expenses of the civil society that they want to exploit or intimidate
\cite{Kleemans:2014tv,enders_sandler_terrorism}. Despite being
inspired by different objectives and modus operandi, theories have
been advanced about organised crime and terrorism laying at the two
extremes of a continuum, in which differences get blurred in a mix of
illegal and violent activities aimed at gaining both economical power
and supremacy over the states
\cite{Makarenko:2010db,Ruggiero:2017dd}. It is not uncommon to see
criminal organisations engaged in terror tactics, or terrorist
networks perpetrating all sorts of criminal activities
\cite{Moro:2015io,Basra:2016cc,Piazza:2017dv,Phillips:2018bw}. Although
empirical evidence about the relevance of a deep crime-terror nexus is
still scarce, it is reasonable to consider that the growth or decline
of COs or TNs are bound to similar dynamics that pertain their
sustainability and attractiveness towards possible recruits.

Modelling crime through the lenses of physics
  \cite{DOrsogna:2015fi} or through computational approaches
  \cite{Groff:2018iz} can provide useful insights to understand the
  dynamics and forces underlying crime development, its relation to
  law enforcement strategies and to evaluate rehabilitation programs
  \cite{Berenji:2014fg,Banerjee:2015jya,Wickes:2018ix,Weisburd:2017cf,Santoprete:2018da}.
  Different approaches have been attempted to model criminal
  behaviour, from abstract dynamical systems models
  \cite{McMillon:2014ke}---possibly including spatial factors
  influencing crime diffusion \cite{Short:2010ib,Zipkin:2014bs}---to
  models following an evolutionary game theory approach, in which
  players display competitive strategies (\eg criminals versus
  punishers) and interact within the context of an adversarial game
  \cite{Short:2010kma,DOrsogna:2013bv,Berenji:2014fg,Perc:2013in,Perc:2015eu}.
  However, despite a raising interest into crime dynamics from a
  complex systems perspective, organised crime and terrorism received
  relatively little attention to
  date~\cite{Mesjasz:2015ch,Szekely:2016iv}.
  Agent-based models have been proposed
  in the context of extortion racket
  \cite{Nardin:2016fv,Szekely:2018bp}. A similar approach has been
  taken to study fundamentalism and radicalisation in the context of
  terrorism
  \cite{daniel2003evolutionary,Udwadia:2006dw,Sandler:2007id,Arce:2009cx},
  but here attention has been directed mainly towards the analysis of
  network properties and the evaluation of disruption interventions
  \cite{Udwadia:2006dw,Keller:2010jo,schwartz2011adaptive,Duijn:2014bo,Li:2015bf,Ren:2019fv}.
  Recruitment to organised crime and terrorism has not been thoroughly
  investigated so far, despite being at the core of the dynamics of
  such groups, and should therefore be one basic ingredient for any
  modelling study that aims at unveiling the complex interactions
  underlying CO/TN dynamics.

In this study, we introduce a stylised model that can be easily
parametrised to represent both the dynamics of COs and of
TNs. Similarly to previous studies
\cite{Short:2010kma,Berenji:2014fg}, we introduce a $N$-person
adversarial game that contrasts agents engaging in illicit activities
and regular \emph{honest} individuals that instead may just lose
out. Differently from previous studies, we differentiate the criminal
population into agents associated into COs or TNs, and individual
agents acting on their own---be they criminals not structured within a
CO, or terrorists not belonging to a TN, hereafter both referred to as
\emph{lone wolves} for brevity. In this way, we can identify the
conditions under which joining a criminal or terrorist organisation is
advantageous. We consider different possible interactions among the
players, taking into account the effects of punishment from
institutions, from the CO/TN itself as well as social control enacted
by the civil society \cite{Sampson:1997bm}. Punishment represents
  a strong driver for positive behaviour, as several theoretical and
  experimental studies demonstrated
  \cite{sigmund2001reward,hauert2007via,han2016synergy,chen2014probabilistic}. Indeed,
  punishment is at the basis of law enforcement strategies, and also
  permeates the world of COs and TNs, making it a fundamental
  mechanism to consider for the study of CO/TN dynamics
  \cite{Short:2010kma,Perc:2013in}.  When focusing on COs, we observe
that under certain conditions criminals contribute to the eradication
of non-organised crime by lone wolves, actually providing a form of
protection to the civil society, as suggested by criminological
theories \cite{Gambetta:93,Varese:2017vp}. When focusing on TNs,
instead, we identify conditions from which TNs coexist in an
equilibrium with lone wolves, effectively benefiting from their
illicit activities without paying related costs
\cite{Feldman:2013be,Spaaij:2010fp}.  In the following, we will
describe the details of the proposed stylised model, and we discuss
the dynamics characterising the above mentioned cases, and beyond.

\section{Methods}

We consider an adversarial game in a multi-agent framework, in which
agents play in small groups formed randomly from a larger well-mixed
population. The latter represents a given community---within a
quartier, a city or a region---that is tight enough to make well-mixed
interactions possible. The small groups within which adversarial games
are played represent instead temporary gatherings that may form and
disband at any time. Agents within the population can take one of the
following roles: (\textbf{i}) \emph{honest citizens} $H$ not committing any
illicit action; (ii) \emph{criminals} $C$ associated to a CO or a TN,
who share the benefits and burden of their actions; and (iii)
\emph{lone wolves} $W$ acting independently from any organisation. The
game is divided into two phases: the \emph{acting stage}, in which
some criminal
activity may take place by a \emph{victimiser} (wolf or criminal) to
the detriment of a subset of agents \emph{victims} within the group;
and the \emph{investigation stage}, in which victimisers may get
punished, either from some state institutions or from other
individuals within the group.  Note that we use the term
\textit{crime} for both criminal and terrorist actions.  After a
certain number of rounds, the evolutionary dynamics take place on the
basis of the payoff cumulated by every agent, modelling the
opportunistic change of strategies within the population. We study
this adversarial game with an analytical mean field model and
complement it with Monte Carlo simulations, as detailed below.

\newcommand{\ZCW}{\langle Z_H, Z_C,Z_W\rangle}
\newcommand{\NCW}[2]{\langle N_H ,N_C, N_W \rangle}
\newcommand{\ZCWp}{\langle Z'_H, Z'_C,Z'_W\rangle_k}
\newcommand{\NCWp}[2]{\langle N'_H ,N'_C, N'_W \rangle_k}

\subsection{Mean Field Evolutionary Model}

We first discuss the details of the proposed adversary game in the
context of a mean field approximation with replicator dynamics
\cite{Traulsen:2006cx}, where we calculate the average payoff that
each agent obtains after an infinite number of rounds played in every
possible configuration of groups in the population.

We consider a well-mixed population of $Z$ individuals.
Within the overall population, each role $k$ is present with a
fraction $x_k=Z_k/Z$, where $\sum Z_k=Z$ and $k=\{H,C,W\}$ stands for
honest, criminal and lone wolf, respectively. We will use the notation
$\ZCW$ to indicate a population formed by $Z_H$ honest citizens, $Z_C$
criminals, and $Z_W$ lone wolves.
From such population, individuals are randomly chosen to form groups
of $N$ individuals, and each group is composed by $N_{k}$ individuals
of each type. Similarly, we refer to a specific group configuration as
$\NCW{}{}$.

\subsubsection{Acting stage} After a group is set up, one agent is
taken randomly from the group to perform its predefined action. The
probability that any given player is chosen from the group is then
$p_1=1/N$, while the probability of choosing a player with role $k$
is $p_k=N_k/N$.  If a honest agent is selected (with probability
$p_H$), nothing happens. Instead, when the selected agent is a
criminal (with probability $p_C$) or a lone wolf (with probability
$p_W$), some criminal action is performed. In the case of lone wolves,
the probability of acting is reduced by a factor
$p'_W=1-\delta\left(1-p_C\right)$, which represents the correlation
between the presence of criminals within the group and the likelihood
that lone wolves take action. For instance, in the terrorism scenario,
lone wolves would act mainly when driven by propaganda from TNs
($\delta=1$), and would otherwise stay quiescent.

Whenever the chosen player is a criminal or a wolf, she will cause a
damage with value $c_k$ in each one of the victims in the group and
obtain a benefit $r_kc_k$ from each one of them. 
If the victimiser is a criminal, honests and wolves are the only
victims and the obtained benefits are shared among all the criminals
within the group (since criminals belong to the same organisation,
they act as a group: they do not damage each other but rather share the
benefits of their actions). The average benefit $b_C$ obtained by
criminals and the average damage $d_{C}$ caused on others can be
computed as follows:
\begin{equation}
  \label{eq:C_benefit_damge}
  b_{C}=\frac{1}{N_C} r_Cc_Cp_C (N-N_C)=r_Cc_C(1-p_C), \qquad d_{C}=c_Cp_C.
\end{equation}
On the other hand, if this victimiser is a wolf, her victims are all
the other members of the group, including other wolves and
criminals. Hence, the average benefit $b_W$ and the average damage
$d_W$ can be computed as follows:
\begin{equation}
  \label{eq:W_benefit_damge}
  b_{W}=r_Wc_W (N-1) p_1p'_W, \qquad d_{W|N'_W}=c_Wp'_W\frac{N'_W}{N},
\end{equation}
where $N'_W$ represents the number of wolves in the group other than
the focal player, and is specified to take into account the
probability that a wolf different from the focal player commits a
crime. Specifically, $N'_W = N_W$ when considering the damage
inflicted by wolves to honests and criminals, while $N'_W = N_W-1$
when considering the damage inflicted by a wolf on the other wolves in
the group.  Finally, we consider the possibility that a fraction
$\tau$ of the benefit obtained by wolves is actually benefiting the
criminals. For instance, a terrorist network would gain in reputation
and power also when the criminal activity is executed by lone wolves
without paying any cost for it.

Overall, from equations~\eqref{eq:C_benefit_damge} and
\eqref{eq:W_benefit_damge}, it is possible to compute the average
payoffs in the acting stage $w^A_k$, as follows:
\begin{align}
  \label{eq:wAH} w^A_{H|\NCW{}{}}&=-d_{C} - d_{W|N_W}\\
  \label{eq:wAC}  w^A_{C|\NCW{}{}}&= b_{C}+\tau b_{W} -  d_{W|N_W}\\
  \label{eq:wAW} w^A_{W|\NCW{}{}}&=(1-\tau) b_{W} - d_{C} - d_{W|N_W-1}
\end{align}
It is possible to notice that honests are only harmed by others in
this stage, while criminals and wolves can get a benefit from illicit
actions, but also suffer from the criminal activities of other
individuals within the group.

\subsubsection{Investigation stage} After each criminal act, an
investigation is conducted. To this end, individuals are chosen from
the group and a control is made on them to ascertain if they committed
a crime.
If the victimiser is found, she
will receive a punishment. We consider three types of investigations
and corresponding punishments:
\begin{itemize}
\item {\bf State}: an investigation performed by a law enforcement
  organisation against any victimiser. The law enforcement
  organisation is not modelled explicitly in the multi-agent
  framework, but as a super-agent. The effects of the corresponding
  investigation are included through the parameter $\beta_S$, which
  represents the level of punishment inflicted to any victimiser and
  is independent from the group/population configuration.
\item {\bf Civil}: social control carried out by the civil
  society---i.e., honest individuals---against any victimiser. When
  successful, this type of investigation leads to the punishment
  $\beta_H$ to be inflicted to the victimisers. In this case, the
  probability of success of such an investigation is proportional to
  the fraction of honest individuals $p_H$.
\item {\bf Criminal}: an investigation performed by the criminal
  organisation against its potential rivals, the wolves. Punishment is
  controlled by the parameter $\beta_C$ and the probability of success
  for the investigation is proportional to the fraction of criminals
  $p_C$. 
\end{itemize}
Note that wolves can receive a higher level of punishment since also
criminals may punish them.  In the case of criminals, instead, we
consider that being part of a CO can lead to the punishment of any
members within the group. The investigated criminal will receive the
full punishment and her partners will receive that punishment reduced
by a factor $\gamma$, since capturing a criminal can lead to capturing
the rest of the criminals involved in the organisation. Overall,
criminals are easier to identify than wolves due to chance only, but
the former have the capacity to punish the latter.  The reductions in
the payoffs that each type of victimiser is obtaining from this stage
$w^I_k$ are, in average, as follows:
\begin{align} 
 \label{eq:wIH} w^I_{H|\NCW{}{}}&=0 \\
 \label{eq:wIC} w^I_{C|\NCW{}{}}&=-\left(\beta_S+ \beta_Hp_H\right) p_C
\left[\gamma p_C + (1-\gamma)p_1\right]\\
 \label{eq:wIW} w^I_{W|\NCW{}{}}&=-\left( \beta_S+ \beta_Hp_H + \beta_Cp_C
\right) p_1^2p'_W
\end{align}
In computing these payoffs, we model the fact that an agent must first
commit a criminal act, which happens with probability $p_1p'_W$ for
wolves and $p_C$ for criminals, and then gets punished upon
investigation, which happens with probability $p_1$ for wolves and
$\left[\gamma p_C + (1-\gamma)p_1\right]$ for criminals, to account
for the collective punishment discussed above.

\subsubsection{Evolutionary dynamics}
 
In order to compute the average payoffs $\omega_k$ that each type of
individual $k$ is obtaining in a given population, we compute the
average payoff that this individual is getting in all the possible
groups she can be part of, considering all the combinations of
remaining $N-1$ individuals in the group.
More specifically, the group of $N$ individuals $\NCW{}{}$ is formed
by the focal player $k$ and a subgroup $\NCWp{}{}$ such as
$N_k=N'_k+1$ and $N_{\tilde k}=N'_{\tilde k}$ for $\tilde{k}\neq k$.
The subgroup is then drawn from the population excluding the focal
player $\ZCWp$ with $Z_k=Z'_k+1$ and $Z_{\tilde k}=Z'_{\tilde k}$ for
$\tilde{k}\neq k$. The likelihood of obtaining $\NCWp{}{}$ drawn from
$\ZCWp$ is given by the multivariate hypergeometric distribution
$\mathcal{H}\left[\NCWp{}{}\right]$:
\begin{equation}
\mathcal{H}\left[\NCWp{}{}\right]=\left \{
\begin{array}{ll}
\frac{\binom{Z'_H}{N'_H}\binom{Z'_C}{N'_C}\binom{Z'_W}{N'_W}}{\binom{Z-1}{N-1}}&\text{if}\;\forall j N'_j\leq\,Z'_j \\
0 & \text{otherwise}\\
\end{array}
\right.
\end{equation}

The average payoff can be then computed starting from equations
\ref{eq:wAH}--\ref{eq:wIW}, and weighting the payoff obtained in each
subgroup with $\mathcal{H}$:
\begin{equation} 
\omega_k = \sum_{N'_C=0}^{N-1}\sum_{N'_W=0}^{N-N'_C-1} \mathcal{H}\left[\NCWp{}{}\right] \left(w^A_{k|\NCW{}{}} + w^I_{k|\NCW{}{}}\right),
\label{eq:payoff_f}
\end{equation}

We assume that the dynamics of the system follow the replicator
dynamics equation. For each subpopulation, we compute the direction
and strength of change as follows:
\begin{equation} 
\label{eq:replicator}
\dot{x}_k=x_k(\omega_k-\overline{\omega}),
\end{equation} 
where $\overline{\omega}=\sum_i x_i \omega_i$ is the average payoff including
every individual of the population. 
In order to calculate the most important (or most visited)
configurations of the finite-size population, we assume that in each
evolutionary time-step, only one individual can change state.
From each possible configuration $\ZCW{}{}$,
we calculate the closer next point on the trajectory determined by the
replicator dynamics from eq.~\ref{eq:replicator}.

In this way, we can build the transition matrix among all the possible
configurations of the population (note that to avoid numerical issues,
  we add a small probability of $\mu=10^{-6}$ to every transition, and
  we renormalise all transition probabilities afterwards ensuring a
  correct transition matrix for the Markov chain).
The stationary distribution from the Markov chain represented by this
matrix in then computed. The probabilities in the stationary
distribution represent the importance of each configuration or, in
other words, the time that the system spends in each configuration
point.

\subsection{Monte Carlo Simulations}
We developed a multi-agent simulation with the purpose of validating
the results from mean-field approximations as discussed above. Mean
field approximations and the replicator dynamics are based on a good
estimation of the average payoff, which determines the outcome of the
evolutionary process. In real systems, estimations are noisy and bound
to many factors, such as the group size or the particular settings of
the studied game. It is therefore important to verify the validity of
the analytical results in the light of the available knowledge.

Simulations are implemented following the same stages as discussed
above. Also in this case, a population of $Z$ individuals evolves with
individuals changing their role between honest, organised
criminals and lone wolves. To compute the payoff
$\omega_j$ for each agent $j$, $G$ games are played, and in each game
the population $Z$ is partitioned in $Z/N$ groups that undergo the
acting and investigation stages. Payoffs are assigned to each
individual and cumulated across different games. After each game, the
population is reshuffled and partitioned to generate another set of
data. Once computed an average payoff for each individual over the $G$
games, the evolutionary step takes place by selecting two players at
random and having them change their role probabilistically according
to their relative payoffs. The probability that agent $i$ copies the
role of agent $j$ is computed according to the Fermi function:
\begin{equation}
  \label{eq:1}
  P(k_i\leftarrow k_j) = \frac{1}{e^{-(\omega_j - \omega_i)/T}+ 1}
\end{equation}
where $T$ is a parameter determining the steepness of the sigmoid
function. With a small probability $\mu$, mutations take place
instead, and one randomly selected individual chooses among the three
roles with equal probability.  Simulations have been optimised to
efficiently compute the evolutionary dynamics over a long time, so as
to identify the trajectory of the system.

\section{Results}

The dynamics grasped by the proposed model are determined by the
chosen parameterisation in a non-intuitive way. We exploit the
possibilities offered by the model to represent different contexts by
fixing a number of parameters that characterise the criminal scenario
(\eg the influence $\delta$ played by criminals on the acting
probability of lone wolves), and for each context we study the
importance of the different types of investigation and punishment onto
the population dynamics by varying $\beta_S$, $\beta_H$, and
$\beta_C$. For each combination of these parameters, we study the
dynamics and calculate the basins of attraction of the population
dynamics by looking at the stationary distribution of the different
individual types, therefore identifying when criminals prevail over
lone wolves or vice versa, or when crime remains under control or gets
totally eradicated from society.

\subsection{Organised crime}

We start by exploring a scenario representing the presence of a CO
within our abstract society. In this case, we consider that there is
no favourable interaction between the CO and the lone wolves acting
independently, neither in the probability of committing a crime (\ie
$\delta=0$) nor in the remission of any benefit (\ie $\tau=0$). In
practice, criminals and wolves are in competition: they commit crimes
independently one from the other, harm each-other and keep the benefit
resulting from their criminal action for themselves.  Without loss of
generality, we assume that wolves and criminals produce the same harm
and obtain the same benefit from it, hence $c_W=c_C$ and
$r_W=r_C=1$. When not stated otherwise, we assume that punishment to
criminals other than the investigated one is halved (\ie
$\gamma=0.5$).

\begin{figure}[!t]
	\centering
	\includegraphics[width=\textwidth]{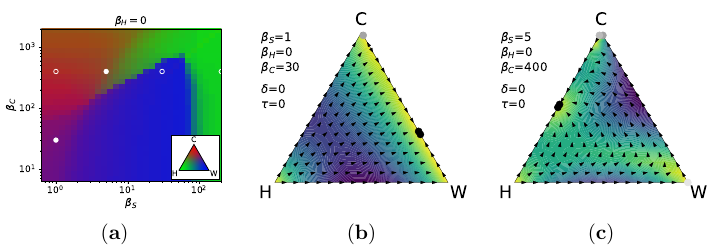}
  \caption{
  Effect of punishment from the state
    ($\beta_S$) and criminal organisation ($\beta_C$) when the civil society
    does not participate to the investigation phase ($\beta_H=0$). 
    (\textbf{a}) Proportion of honest individuals (green), lone wolves (blue) and
    criminals (red) in the stationary distribution for different
    values of $\beta_S$ and $\beta_C$. The inset shows the
    colour-coding corresponding to every point in the simplex
    representing the percentage of each individual type within the
    population.  Circles correspond to representative configurations
    displayed in the accompanying panels (filled circles) or in the
    Supplementary Figure~S1 (empty
    circles). (\textbf{b}-\textbf{c}) Simplex describes the dynamics of the system for
    $\beta_C=400$ and different values of $\beta_S$. Arrows represent
    the direction of change for the three sub-populations, while the
    background colour represents the intensity of change---the darker
    the stronger. Filled circles indicate the most visited states: the
    grey scale represents the corresponding probability in the
    stationary distribution---the darker the higher---normalised
    according to the most visited state and using a minimum threshold
    of 10\%. Parameters of the model for this figure: $\gamma=0.5$,
    $N=10$, $c_W=c_C=r_W=r_C=1$, $Z=50$.}
  \label{fig:bS-bC_bH0}
\end{figure}

We first consider the case in which the civil society is not
contributing to investigations and punishment (\ie $\beta_H=0$). In
Figure~\ref{fig:bS-bC_bH0}, we represent the proportion of each
individual type in the stationary distribution for a wide range of
values for $\beta_S$ and $\beta_C$, as well as the system dynamics for
representative combinations. The stationary distributions reveal the
presence of different possible regimes, and indicate that punishments
from the state organisation and from criminals interact in a
non-trivial way (see Figure~\ref{fig:bS-bC_bH0}a). As expected, when
punishment from criminals is relatively small ($\beta_C < 50$), wolves
take over the population, unless the punishment $\beta_S$ is also very
small. In correspondence of a weak state, there exist a region in
which criminals and wolves coexist at the expenses of the civil
society (see Figure~\ref{fig:bS-bC_bH0}b), and also a region where the
CO proliferates by increasing punishment against lone wolves (as shown
in Figure~\ref{fig:bS-bC_bH0}c, where $\beta_C=400$). In this latter
condition, the CO practically takes over the role of the state in
punishing criminal acts carried out by wolves, providing a form of
protection \cite{Gambetta:93}. The model predicts
that, in correspondence of a low punishment from the state
organisation, criminals and honests can coexist in the society. With
increasing levels of punishment from the state organisation, the CO
gradually disappears to the benefit of the civil society. However, an
undersized CO also has a low control potential against lone wolves,
leaving room to their proliferation:
when a specific value of $\beta_S$ is reached (see for instance
$\beta_C=400$ and $\beta_S=200$ in Figure~\ref{fig:bS-bC_bH0}), a
phase transition takes place and wolves take over the entire
population. Indeed, $\beta_S$ is high enough to undermine the power of
criminals but not to punish efficiently wolves. Without criminals able
to control wolves and a state not strong enough to do it by itself, the
population is at mercy of wolves.
Only with a sufficiently strong state punishment (\eg $\beta_S>100$)
crime can get completely eradicated (see Figure~\ref{fig:bS-bC_bH0}a
and the Supplementary Figure~S1).
The fact that wolves can have an advantage over criminals for medium
values of $\beta_S$ finds its explanation in the way in which
investigations are modelled, which make the identification of a
criminal group much more likely than a single lone wolf.  This is a
simplification introduced by the model, which however shows the
interesting effects introduced by a differential probability of being
caught. Additionally, punishment is not limited to the investigated
criminal, but also affect the other criminals in the group via the
factor $\gamma$. Indeed, the CO dominates in a large part of the
parameter space with low values of $\gamma$, and conversely is less
powerful when $\gamma$ is high (see the Supplementary Figure
S2).

\begin{figure}[!t]
	\centering
	\includegraphics[width=\textwidth]{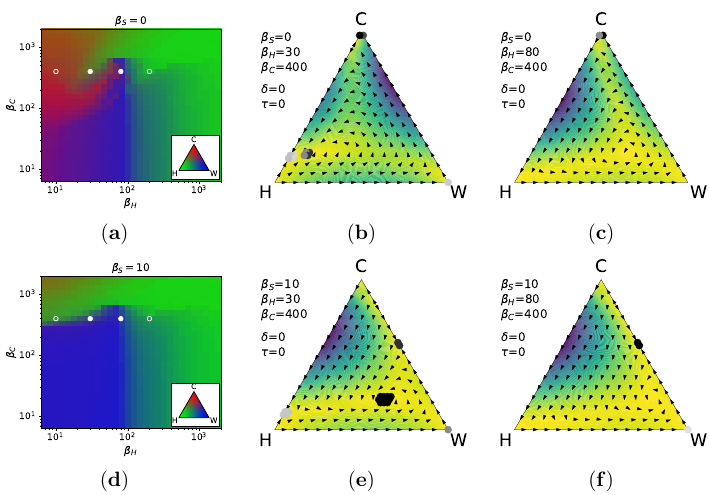}
        \caption{ Effect of punishment from honest individuals
          ($\beta_H$) and criminal organisation ($\beta_C$) when the
          state is absent ($\beta_S=0$, top row) or when it is present
          but weak ($\beta_S=10$, bottom row).
          (\textbf{a},\textbf{d}) Stationary distribution for
          different values of $\beta_H$ and $\beta_C$. Circles
          represent relevant configurations shown in the side panels,
          or in supplementary Figure
          \ref{fig:bH-bC_bS10_supp}. (\textbf{b},\textbf{c},\textbf{e},\textbf{f})
          Dynamics of the model under specific parametrisations. See
          Figure~\ref{fig:bS-bC_bH0} for additional details.
          Parameters of the model: $\gamma=0.5$, $N=10$,
          $c_W=c_C=r_W=r_C=1$, $Z=50$. }
  \label{fig:bH-bC_bS0}
\end{figure}

Slightly different dynamics are observable when punishment comes also
from the civil society (\ie $\beta_H > 0$). When the state
organisation is absent ($\beta_S=0$, see Figure~\ref{fig:bH-bC_bS0}\textbf{a}),
crime control falls on the shoulder of honest individuals, and their
effectiveness is proportional to the size of the honest
population. Hence, larger punishment values are required to produce a
similar effect as with the state organisation.  Also in such
conditions, criminals prevail for low values of punishment $\beta_H$,
and are then gradually replaced by wolves when the punishment
increases. Punishment from criminals to lone wolves provides an
advantage to the former as long as $\beta_H$ remains sufficiently
small (see Figure~\ref{fig:bH-bC_bS0}\textbf{b}).  However, instead of having a
clear phase transition to the domination of wolves, now spirals in the
population dynamics appear (see Figure~\ref{fig:bH-bC_bS0}\textbf{c}). In the
presence of many honest individuals, criminals are those that lose out
first, as they are easily victim of investigations. When the fraction
of criminals is low, they do not sufficiently contribute to control
the wolves, which can therefore dominate. However, as soon as honest
are not enough to be exploited and to keep criminals under control,
the residual punishment from criminals leads again to the recovery of
the CO.  Figure \ref{fig:bH-bC_bS0}\textbf{c} shows that these dynamics are
captured by a slow repelling spiral for $\beta_S=0$, eventually
leading to the dominance of organised crime. Depending on the specific
value of $\beta_H$ and $\beta_C$, the final state can change favouring
the one or the other group (see also the Supplementary
Figure~S3\textbf{a}-\textbf{b}). 
The combined action of punishment from the state organisation and from
the civil society has a strong effect on the subsistence of the CO
(see Figure~\ref{fig:bH-bC_bS0}\textbf{d}, where $\beta_S=10$). In this case,
lone wolves dominate in a large region of the parameter space. We can
observe again spiralling dynamics (see Figure~\ref{fig:bH-bC_bS0}\textbf{e}),
which however tend to converge to a mixed equilibrium with many wolves
and few honests and criminals.  With increasing punishment from the
civil society, the equilibrium shifts in favour of wolves first, and
honest individuals later (see Figure~\ref{fig:bH-bC_bS0}\textbf{f} and the
Supplementary Figure~S3).  We have also
observed that criminals can coexist in a stable way with both wolves
and honest. However wolves and honests are not able to coexist and
maintain an equilibrium: if both are present in the population, they
also need criminals for a stable coexistence.

As already mentioned, the parameter $\gamma$ influences the success of
the CO over lone wolves: the lower the punishment towards
co-offenders, the higher the power of the criminal organisation with
respect to lone wolves (see Supplementary Figure~S2).
The size of the groups $N$ also significantly affects the population
dynamics: the larger the groups, the larger the benefit for
victimisers, especially for wolves as shown in the Supplementary
Figure~S4. In order to obtain a better insight, one can
deduce how much punishment is necessary for eradicating victimisers
under the limit of $N=Z$ and when only two types of individuals are in
the population (note that most of the resting points are found on the
borders of the simplexes, hence justifying this assumption). Comparing
the payoffs $\omega^A_k+\omega^I_k$ from
Eqs.~\ref{eq:wAH}--\ref{eq:wIW} of each two pair of types of actors
(and assuming for simplicity that $r_C=r_W=1$, $c_C=c_W=c$, and
$\delta=\tau=0$), we obtain that:
\begin{itemize}
\item If $x_C=0$, honest individuals defeat wolves if $\beta'>N^2$
\item If $x_W=0$, honest individuals overcome criminals if
  $\beta'>(g\, p_C)^{-1}$
\item If $x_H=0$, criminals defeat wolves if $\beta_C>(N^2g\, p_C-1)\beta_S$	
\end{itemize}
where $\beta'=(\beta_S+\beta_Hx_H)\,c^{-1}$ and
$g=(1-\gamma)p_1+\gamma p_C$. In a population with only wolves and
honest individuals, the punishment $\beta'$ required for the latter to
overcome the former is proportional to $N^2$, confirming that larger
groups provide a benefit to wolves. If instead of wolves the
victimisers are only criminals, that punishment depends on how
trackable are criminals, as determined by the parameter $\gamma$: if
from one criminal it is easy to catch the others
($\gamma\approx 1 \rightarrow g\approx p_C$), the punishment
$\beta'$---required to observe honest individuals prevailing on
criminals---decreases with the square of the fraction of criminals
($\beta'>p_C^{-2}$), whereas in the opposite case
($\gamma\approx 0 \rightarrow g\approx p_1$) this punishment required
decreases with the fraction of criminals but increases with the size
of the population ($\beta'>N\,p_C^{-1}$). This illustrates why both
victimisers increase their power in bigger groups, but wolves do that
in a greater way. Finally, in the competition between wolves and
criminals, the latter gets a disadvantage under strong punishment from
the state organisation, as well as when $\gamma$ is high: for
$\gamma\approx 1$, the punishment that criminals have to inflict to
wolves is $\beta_C>(N_C^2-1)\beta_S$, scaling quadratically with the
size of the CO, so as to compensate the costs paid from the
investigation stage.

To understand under what conditions the criminal strategy is viable,
we tested different parametrisation for the system by varying the
level of harm inflicted by criminals and wolves during the acting
stage, and the relative benefit they obtain from it (see Supplementary
Figure~S5). In all cases, the stationary distribution is
similar to the main cases described above, although scaled in favour
of the one or the other type of victimiser, as expected by the
relative strength provided by different costs-to-benefit
ratios. Overall, following a CO pays off both when the harm inflicted
is higher, as well as when the resulting reward is bigger than those
of wolves. This means that, under the considered conditions, the CO
needs to get some advantage in terms of professionalisation of the
criminal activities in order to be sustainable.

\begin{figure}
  \centering
	\includegraphics[width=\textwidth]{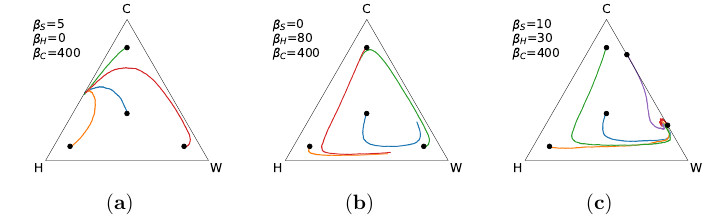}
  \caption{
  Dynamics resulting from Monte Carlo simulations for different values
    of $\beta_S$, $\beta_H$, and $\beta_C$. Different colours show the
    dynamics from different initial conditions (marked here with a black
    filled circle) averaged over 500 realisations. Parameters of the model:
    $\delta=0$, $\tau=0$, $\gamma=0.5$, $N=10$, $c_W=c_C=r_W=r_C=1$.}
  \label{fig:sims}
\end{figure}

A set of Monte Carlo simulations were performed ratifying the results
obtained using mean-field approximation. In Figure~\ref{fig:sims},
three different conditions are shown, respectively corresponding to
Figure~\ref{fig:bS-bC_bH0}\textbf{c}, \ref{fig:bH-bC_bS0}\textbf{c} and
\ref{fig:bH-bC_bS0}\textbf{e}. Each trajectory has been obtained by averaging
over 500 independent runs, each run lasting 50000 iterations. At each
iteration, $G=100$ games have been performed to compute the average
payoff of each individual within the population. The average
trajectories shown in Figure~\ref{fig:sims} closely correspond to the
theoretical predictions, although the spiral center in
Figure~\ref{fig:sims}\textbf{c} is shifted to the right.

\subsection{Terrorist networks}

To model the terrorist scenario, we assume that criminals are
terrorists that belong to an organised network, while wolves are
terrorists that act on their own. We use the same names to refer to
them in spite of the different roles they represent. 
With respect to the organised crime case, there are three important
aspects that need to be taken into account. First, criminals organised
in a TN have aligned interests with lone wolves. Both are willing to
threaten and destabilise society, and are therefore not in
competition. To model this aspect, we remove the direct punishment
from criminals to wolves ($\beta_C=0$).
Second, the TN can directly benefit from criminal
activities perpetrated by lone wolves, by gaining in power and in
reputation at the expenses of the individuals that are directly
involved in the terrorist act, who sometimes even sacrifice their
lives for the cause. We model similar aspects of the terrorist
interaction as a transfer of benefit from wolves to criminals by a
fraction $\tau\in[0,1]$, where a value of 1 implies a full transfer of
benefit.
Third, we consider the possibility that the TN actively promotes
actions by lone wolves through their propaganda against the
state. This is an effective strategy that can be modelled by linking
the probability that lone wolves commit a terrorist act to the size of
the TN: the larger the network, the stronger the propaganda, the more
likely lone wolves act.  We model this through the parameter $\delta$
linking the likelihood of wolves' action to the fraction of criminals
in the group.

\begin{figure}[!t]
  \centering
	\includegraphics[width=\textwidth]{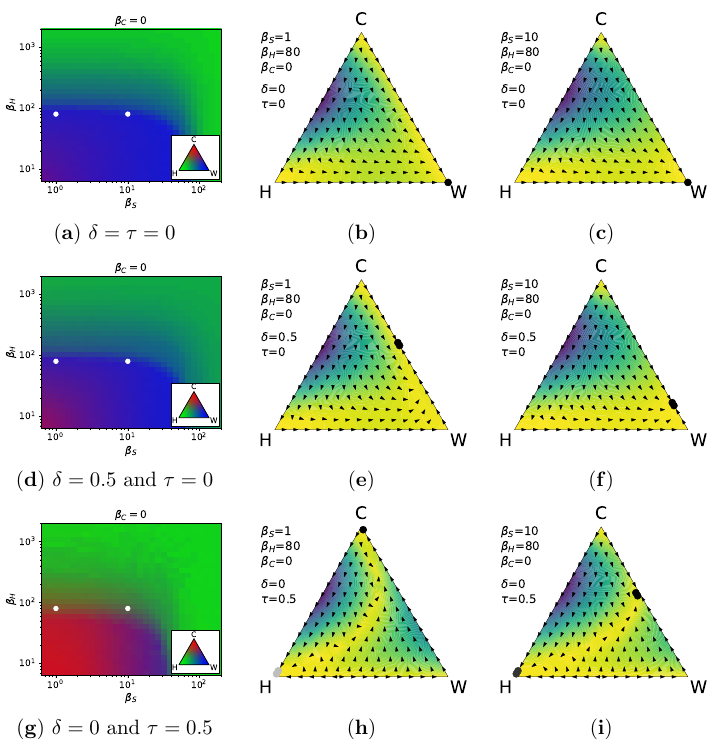}
  \caption{Effect of the propaganda ($\delta$) and transfer of benefit
    ($\tau$) in modelling the dynamics of TNs. We consider a baseline
    case in which neither of the proposed mechanisms are present
    ($\delta=0$ and $\tau=0$, first row), the case in which only
    propaganda is used ($\delta=0.5$ and $\tau=0$, second row), and
    the case in which only the transfer of benefit is considered
    ($\delta=0$ and $\tau=0.5$, third row).  The combined effects of
    both mechanisms is shown in the Supplementary
    Figure~S6. (\textbf{a},\textbf{d},\textbf{g}) Each panel shows the
    stationary distribution of the different roles, similar to
    Figure~\ref{fig:bS-bC_bH0}\textbf{a}, but for $\beta_C=0$ and different
    values of $\beta_H$ and $\beta_S$. Filled circles represent
    relevant configurations displayed. (\textbf{b},\textbf{c},\textbf{e},\textbf{f},\textbf{h},\textbf{i}) Dynamics of the
    model under relevant parametrisations. See
    Supplementary Figure~\ref{fig:bS-bC_bH0} for further details.}
	\label{fig:TNdeltatau}
\end{figure}

The effects of the parameters $\tau$ and $\delta$ on the stationary
distribution of roles within the population reveals that propaganda
alone does not determine the dominance of the TN within the population
(see Figure~\ref{fig:TNdeltatau}). More specifically, we can observe
that, when there is no propaganda and no transfer of benefit, the TN
persists only for very low values of punishment $\beta_S$ from the
state organisation and $\beta_H$ from the civil society (see
Figure~\ref{fig:TNdeltatau}\textbf{a}). Otherwise, wolves take over the entire
population (see for instance the dynamics displayed in
Figure~\ref{fig:TNdeltatau}\textbf{b}-\textbf{c}), unless punishment is high enough
($\beta_S>100$ or $\beta_H>1000$ ).

\section{Conclusions}

We have presented an evolutionary model that grasps interesting
  dynamics related to the proliferation of organised crime and
  terrorism in a population. The model accounts for offences
  perpetrated by criminals, individually or organised into criminal
  organisations or terrorist networks, at the expenses of other agents
  in the population. 
  Different forms of punishment are considered to control the spread
  of crime into the modelled society, including social control by
  institutions and the civil society as well as punishment from
  organised criminals towards lone wolves.

A close look to the results presented in this study provides support
to criminological theories about the development of organised crime
and terrorist networks. Related to organised crime, our model predicts
the establishment of a CO as an entity capable of providing protection
to the population \cite{Gambetta:93,Varese:2017vp}, especially when
they face a relatively weak state and when the civil society does not
participate to punishment. The CO in this case takes the role of
governance, controlling the criminal acts from wolves to impose their
own rules. When instead honest citizens oppose themselves to the CO,
an equilibrium is unlikely and cycles may be observed in which
criminal activities grow and shrink in response to the reaction from
the civil society. Similar complex patterns are found also in other
modelling approaches to studying crime \cite{Perc:2013in}, however in
this case they are the result of dominance among different strategies
that are made possible by frequency-dependent punishment, which
introduces non-linearities that are relevant for the emergence of
complex dynamical patterns. 

Related to terror networks, we observed how the role of propaganda is
central in promoting a stable equilibrium between the TN and the lone
wolves, more than a possible transfer of benefit from wolves to the
TN. Propaganda is recognised as a central aspect for terrorism, and is
found in all its manifestations, especially recently with the strong
exploitation of the Internet and of social media technologies
\cite{Conway:2006kd}. The latter provide a rather cheap mean of
publicising terror strategies, although their effect in effectively
promoting radicalisation and terrorist acts still needs to be
ascertained \cite{Conway:2017kf}. The model predicts that, when
propaganda is successful, the TN can remain relatively small in size,
but the terror goal is anyway reached thanks to the action of lone
wolves. There is instead less evidence of a direct transfer of benefit
from wolves to the TN other than a return in reputation and strength
of the TN that claims responsibility on the attack, which however can
be framed in a sort of competition between the terrorist organisation
and the lone wolves \cite{Alakoc:2015ja}. Ultimately, criminology
research confirms that any TN needs lone wolves either as autonomous
cells that can perpetrate terror acts on their own, or as possible
recruits to increase their dimension and power
\cite{Spaaij:2010fp,Feldman:2013be}. The proposed model can help in
grounding different conceptual frameworks into tangible social
dynamics.

Overall, despite being conceptually very simple, the proposed model
grasps interesting patterns related to the development of CO and TN
when embedded in a society in which also non-organised criminal
activities are present.  Real-world scenarios are clearly more
  complex than what pictured here, as they can involve both
  competition and collaboration between different criminal
  organisations, which can behave in a range of different ways against
  non-organised criminals, at times punishing, ignoring or promoting
  their activities. For instance, in our model, we do not consider the
  possibility of retaliation, which has been found to be relevant in
  other studies on crime dynamics \cite{Short:2010kma,DOrsogna:2015fi}
  and can be related to theoretical studies of antisocial punishment
  \cite{herrmann2008antisocial,rand2010anti,hilbe2012emergence,szolnoki2017second}. Nevertheless,
  the dynamics we observe are already rich enough to provide useful
  accounts on the underlying CO/TN dynamics, which could be matched
  with real-world instances.  In this respect, an interesting
  possibility to study the dynamics of organised crime with respect to
  non-organised criminals is looking at the transplantation of a CO,
  whereby new territories not occupied by other COs are colonised by
  branches of organisations elsewhere very
  powerful~\cite{Varese:2011cs}. Similar conditions offer an
  interesting opportunity to contrast the prediction from the proposed
  model to real-world observations, pointing to the need of collecting
  precise data about the spread of COs in correspondence to
  transplantation attempts.

The complex dynamical patterns that emerge from the study point to the
need to take into account the social context in which certain criminal
behaviour are observed.  Indeed, more than individual predisposition
to criminal activities, the theory of \emph{social opportunity
	structure} \cite{Kleemans:2008fq} postulates that involvement into
criminal activities is strongly determined by social contacts and
contingent opportunities that can become available at any point in the
life of an individual. The evolutionary perspective that we take in
this paper follows from the explanations provided by the social
opportunity structure, as the mutation towards criminal activities is
determined by the opportunities that are given by higher potential
payoffs. To further build on this theory, models can be developed in
which social ties are explicitly taken into account, for instance by
having agents interact on heterogeneous networks, possibly changing
over time to adapt to social contingencies, hence strengthening or
loosing ties \cite{schwartz2011adaptive}. Interesting perspectives can
be given by studying the evolutionary dynamics on multilayer networks
\cite{Kivela:2014dm,Boccaletti:2014bza}, which can grasp the existence
of different types of relations between agents (family ties,
friendship, work ties and so forth). In this way, it could be possible
to include more detailed mechanisms for recruitment and radicalisation
of individuals, further testing theories related to the social
opportunity structure that accompanies the evolution of organised
crime.

\section*{Acknowledgments}
The authors wish to thank Francesco Calderoni and Giulia Andrighetto for their insightful comments. This work was partially supported by the project DICE (FP7 Marie Curie Career Integration Grant, GA: 631297) and the Horizon 2020 Project PROTON ``Modelling the Processes leading to Organised crime and Terrorist Networks'' (GA: 699824).
LAMV also acknowledges the support of the European Research Consortium for Informatics and Mathematics through an Alain Bensoussan Fellowship.
\section*{Author contributions}
LAMV and VT designed the research, analysed the results, and wrote the manuscript. The mean-field model was implemented by LAMV and the Monte Carlo simulations by VD.
\section*{Additional Information}
The authors declare no competing interests.

\bibliographystyle{naturemag}

\end{document}